\documentstyle[aps]{revtex}

\newcommand{\str}{\mathop{\rm str}\nolimits}
\newcommand{\tr}{\mathop{\rm tr}\nolimits}

\renewcommand{\vr}{{\bf r}}
\newcommand{\vp}{{\bf p}}
\newcommand{\vn}{{\bf n}}
\newcommand{\vq}{{\bf q}}
\newcommand{\vnabla}{{\bf \nabla}}
\newcommand{\vM}{{\bf M}}
\newcommand{\vB}{{\bf B }}
\newcommand{\vJ}{{\bf J}}

\begin{document}

\author{ B.A.Muzykantskii and D.E.Khmelnitskii}
\address{Cavendish Laboratory, University of Cambridge, Madingley Road,
Cambridge, CB3 0HE, UK \\ and L.D.Landau Institute for Theoretical
Physics, Moscow, Russia}
\title{Effective Action in Theory of
Quasi-Ballistic Disordered Conductors}

\date{May 30, 1995}

\maketitle
%%%%%%%%%%%%%%%%%%%%%%%%%%%%%%%%%%%%%%%%%%%%%%%%%%%%%%%%%%%%%%%%
\begin{abstract}
We suggest an effective field theory for disorderd conductors, which
describes quantum kinetics of ballistically propagating electrons.
This theory contains non-linear $ $ $\sigma$-model \cite{Efetov}
as its long wave limit.
\end{abstract}

\bigskip
\large

$\bf 1$. The non-linear $\sigma$-model is proven to be a useful tool
in the description of various properties of disordered conductors. Any
property, such as conductivity, averaged over different realizations of
the random potential can be presented in this model as
a statistical average with the free energy
\begin{equation}
  F=\frac{\pi \nu}{8} \int d \vr \str [ D (\nabla Q)^2 + 2 i \omega
  \Lambda Q ], \qquad Z =\int\limits_{Q^2=1} \!\!\! {\cal D} Q e^{-F}
        \label{Ef}
\end{equation}
The functional integral is taken over the $8 \times 8$
super-matrix $Q(\vr)$ which is subjected to the constraint $Q^2 =1$.
Here and below we use the super-matrix version \cite {Efetov}
of the nonlinear $\sigma$ model.

This discription is valid under the following two conditions:
\begin{enumerate}
     \item The Fermi wave length $\lambda_F = \hbar /p_F $ is much smaller
          than the mean free path $l$, i.e. $p_F l /\hbar \gg 1$.

     \item The typical wave vector $q$ of the super-matrix fluctuations is
        smaller than $1/l$, i.e. $ q l \ll 1$.
\end{enumerate}

These conditions mean that (i) the semi-classical description is
applicable to the electrons with the Fermi energy, and (ii) their
motion is described by the diffusion equation.  There are physical
situations when the condition (i) is fulfilled, while the
condition (ii) is not and electrons propagate ballistically. This
happens, for example in a metallic grain with a diffusive boundary
scattering if the bulk mean free path $l$ is much larger than the grain
size $L$, i.e. $l \gg L$.

In this letter we present a generalized version of the model (\ref{Ef})
whose validity is no longer restricted by condition (ii). The generalized
partition function correctly accounts for the fluctuations with wave vectors
$ q \sim 1/l $ and therefore can be used for the description of systems with
ballistic electron motion.

We begin with a general expression for the free energy which is
obtained after averaging over the random potential, the
Hubbard-Stratonovich decomposition of the quartic form and integration
over the electron degrees of freedom (see \cite {Efetov} for details and
notations).
\begin{mathletters}
\label{log}
\begin{eqnarray}
  F &=& -\frac{1}{2} \str \ln [-i \hat K] + \frac{\pi \nu}{8 \tau} \int
  \str Q^2(\vr) d \vr, \quad Z = \int {\cal D} Q e^{-F},
 \label{log1}
\\
  \hat K &=& E- \hat H_0 +\frac{\omega}{2} \Lambda +
 \frac{i}{2 \tau} Q, \quad
        \hat H_0 = \frac{(-i\hbar \vnabla)^2}{2m}
\end{eqnarray}
\end{mathletters}
This expression appears at a preliminary stage in the derivation of
Eq.~(\ref{Ef}) and the supermatrix  $Q$ is not yet restricted by the
constraint $Q^2=1$.

Equation~(\ref {log1}), in principle, could have served as a required
generalisation of the free energy (\ref {Ef}).  However, it is too
detailed being valid for the super-matrices $Q$ fluctuating with
arbitrary wave vectors $\vq$. It will be simplified in order to describe
the small $\vq$ fluctuations only ($q \ll p_F/\hbar$).  The first step in
the simplification is the same as in the derivation of the quantum
kinetic equation in the Keldysh approach (see, for example, \cite
{Schmid}).

%%%%%%%%%%%%%%%%%%%%%%%%%%%%%%%%%%%%%%%%%%%%%%%%%%%%%%%%%%%%%%%%%%%%%%%%%%
$\bf 2$. The Green function $G(\vr, \vr'|Q)$ of the operator
$\hat K$ obeys the equations
\begin{mathletters}
\label{Dyson}
\begin{eqnarray}
  \left[ E - \hat H_0 (\vr) +\frac{\omega}{2} \Lambda + \frac{i}{2 \tau}
    Q(\vr) \right] G(\vr,\vr'|Q) & = & i \delta ( \vr-\vr')
   \label{Dyson1}
\\
  \left[
    E- \hat H_0 (\vr')
  \right] G(\vr, \vr'|Q) +
  G(\vr, \vr'|Q)  \left[
                    \frac{\omega}{2} \Lambda + \frac{i}{2 \tau} Q(\vr')
     \right]  &=& i \delta (\vr-\vr')
        \label{Dyson2}
\end{eqnarray}
\end{mathletters}
Subtracting Eq.~(\ref {Dyson2}) from Eq.~(\ref {Dyson1}) and going to
the Wigner representation
\begin{equation}
  G(\vr, \vr') =\int (d \vp) \, \tilde{G} (\frac{\vr +\vr'}{2},\vp) \,
    e^{i \vp (\vr- \vr')}
        \label{Wigner}
\end{equation}
we can find after the integration over the modulus of the momentum $ \vp $
an equation for
\begin{equation}
  g_\vn (\vr) = \frac{1}{\pi} \int d \xi \tilde {G}(\vr,\vn
  \frac{\xi}{v_F}) , \qquad \vn^2 = 1.
        \label{Eilen1}
\end{equation}
This equation can be presented in the form
\begin{equation}
        2v_F \vn \frac{\partial g_{\vn} (\vr)}{\partial \vr} =
        \left[i\omega \Lambda - \frac{Q}{\tau}, g_\vn \right],
        \label{Eilen2}
\end{equation}
which resembles the  quantum kinetic equation in the Eilenberger form
\cite {Eilenberger}.
The matrix $ g_\vn (\vr) $ in this equation has
the meaning of distribution function at a coordinate $ \vr $ and
momentum $\vp = \vn \cdot p_F$.

Being linear, Eq.~(\ref {Eilen2})
does not define $g_\vn$ uniquely and must be
supplied with the normalisation condition \cite {Schmid}
\begin{equation}
        g_\vn^2 = 1 ; \qquad \tr g_\vn = 0.
        \label{Norm}
\end{equation}
The matrix $Q(\vr)$ is invariant with respect to the charge conjugation
\begin{equation}
        \bar{Q} \equiv CQ^T C^T = Q,
        \label{bar}
\end{equation}
where $\hat{C}$ is a certain matrix (see \cite{Efetov}), $C^T C = 1$. Taking
the charge conjugate of Eq~(\ref{Dyson1}) and using  Eq~(\ref{bar}), we see
that $\bar{G}(\vr, \vr')$ obeys Eq~(\ref{Dyson2}). Therefore
\begin{equation}
  \bar{G}({\vr, \vr'}) = G(\vr',\vr), \quad
  \bar{\tilde{G}}(\vr, \vp) = G(\vr, -\vp), \quad
  \bar{g}_\vn (\vr) = g_{ -\vn} (\vr).
        \label{bar1}
\end{equation}
Thus, Eq.~(\ref {Eilen2}) with the normalisation condition (\ref{Norm})
and the symmetries (\ref{bar1}) is a long wave
limit of Eqs.~(\ref {Dyson}).
Our goal is to perform analogous simplification of the free energy
(\ref{log1}).

%%%%%%%%%%%%%%%%%%%%%%%%%%%%%%%%%%%%%%%%%%%%%%%%%%%%%%%%%%%%%%%%%%%%
%%%%%%%%%%%%%%%%%%%%%%%%%%%%%%%%%%%%%%%%%%%%%%%%%%%%%%%%%%%%%%%%%%%%%%

$\bf 3$.  An intermediate step is finding a functional $ \Phi $,
which reaches its extrema for solutions of  Eq.~(\ref
{Eilen2}). This equation resembles the equation of motion of a magnetic
moment $ \vM $ in external magnetic field $ \vB $ :
 \begin{equation}
  \frac{\partial \vM}{\partial t} = [\vM \times \vB], \qquad \vM^2 = 1.
\label{Bloch}
\end{equation}
The action for this problem has the form (see, for instance, \cite{Klauder})
 \begin{equation}
   {\cal A}= \int_0^t dt' \vB \vM(t') + \int_0^t dt'\int_0^1 du
   \tilde \vM \cdot \left [\frac{\partial \tilde \vM}{\partial t} \times
     \frac{\partial \tilde \vM }{\partial u}\right ],
\label{Klaud}
\end{equation}
where the function $\tilde \vM (t,u)$ is introduced as
\begin{equation}
  \tilde \vM (t,0) = \vM_0; \qquad \tilde \vM (t,1)=\vM (t).
        \label{cond}
\end{equation}
The second term in Eq (\ref {Klaud}) does not depend upon the choice of
$\vM_0$ and values of $\tilde \vM (t,u)$ for $0<u<1$, provided $\vM (0)=
\vM(t)$.

Following this analogy we present $ \Phi $ in the form
\begin{mathletters}
\label{Vic}
\begin{eqnarray}
  &&{\Phi} = \int d \vr \str \left[ (\frac{1}{\tau} Q(\vr) - i \omega
    \Lambda) \langle g (\vr) \rangle \right] + \frac{v_F}{2} {\cal
    W}\{g_\vn\},
\label{Vic1}
\\
&&\langle  g(\vr) \rangle =
\int \frac{d \Omega_\vn}{4 \pi} g_\vn (\vr),
\label{Vic2}
\\
&&{\cal W}\{g_\vn\} = \int d \vr \int \frac{d
  \Omega_\vn}{4 \pi} \int_0^1 du \str \tilde g_\vn (\vr,u) \left[\frac{\partial
\tilde
  g_\vn}{\partial u},
  \vn \frac{\partial \tilde g_\vn} {\partial \vr} \right],
\label{W}
\\
&&\tilde g_\vn (\vr,0) = \Lambda;\; \qquad \tilde g_\vn(\vr,1)= g_\vn(\vr).
\label{Vic3}
\end{eqnarray}
\end{mathletters}
The functional derivative $\delta {\Phi} / \delta g_{\vn}$ must be
taken with constraint (\ref{Norm}) which guaranties that $ g_\vn
\delta g_\vn + \delta g_\vn g_\vn =0 $ and an arbitrary
variation  $\delta g_\vn$ has the form
  $\delta g_\vn = \left[ g_\vn,  a_\vn \right]$.
As a result
\begin{mathletters}
\label{delta}
\begin{equation}
  \label{deltaF}
  \delta {\Phi} = \int d \vr \int \frac{d \Omega_\vn}{4 \pi}
\str \left(
   \left[ \frac{1}{\tau} Q(\vr) - i \omega \Lambda, g_\vn \right] a_\vn
 \right) + \frac{v_F}{2} \delta {\cal W},
\end{equation}
where
\begin{equation}
\label{deltaW}
\delta {\cal W} = 4 \int d \vr \int \frac{d \Omega_\vn}{4 \pi}
\str \left(
 \vn \frac{\partial g_\vn}{\partial \vr} a_\vn \right).
\end{equation}
\end{mathletters}
Thus, Eq.~(\ref{Vic1}) gives the required functional.

%%%%%%%%%%%%%%%%%%%%%%%%
%%%%%%%%%%%%%%%%%%%%%%%%%%%%%
$\bf 4$. Now we are prepared to show that in the limit $l \gg \lambda_F$ the
partition function (\ref{log1}) reduces to the form
\begin{mathletters}
\label{BigVic}
\begin{eqnarray}
 &&Z=\int\limits_{g_\vn^2=1} \!\!\! {\cal D} g_\vn (\vr) e^{-F},
\label{BigVic1}
\\
&&F = \frac{\pi
    \nu}{4} \left[ \int d \vr  \str \left\{
    i\omega \Lambda \langle g (\vr) \rangle  - \frac{1}{2 \tau}
\langle g(\vr)\rangle^2 \right\} -
\frac{v_F}{2} \cdot {\cal W} \{ g_\vn \} \right],
\label{BigVic2}
\\
&&{\cal W}\{g_\vn\} = \int d \vr \int \frac{d
  \Omega_\vn}{4 \pi} \int_0^1 du \str \tilde g_\vn (\vr,u)
  \left[\frac{\partial \tilde g_\vn}{\partial u},
  \vn \frac{\partial \tilde g_\vn} {\partial \vr} \right].
\label{BigVic3}
\end{eqnarray}
\end{mathletters}
Indeed, the following indentity is valid
\begin{equation}
  Z_1\{Q\} \equiv \exp \left[\frac{1}{2} \str \ln(-i \hat K )\right] =
  \int\limits_{g_\vn^2=1} \!\!\! {\cal D} g_\vn (\vr) \exp \left[
    \frac{\pi \nu}{4} {\Phi} \right] \equiv Z_2 \{Q\}.
        \label{ind3}
\end{equation}
The free energy $\pi \nu {\Phi}/4 $ in the partition function $Z_2$ has
a deep minimum for $g_\vn (\vr)$ equal to $ g_\vn ^{(0)} (\vr |Q)$ which
is the solution of Eq.~(\ref{Eilen2}). With the saddle point precision
\begin{equation}
  \frac{\delta Z_2 \{Q\}}{\delta Q (\vr)} =
 \frac{\pi \nu}{4 \tau}\int \langle g(\vr) \rangle \exp \left[
    - \frac{\pi \nu}{4 \tau} {\Phi} \right] {\cal D} g_\vn =
\frac{\pi \nu }{4 \tau}  \langle g_\vn ^{(0)} \rangle \cdot
Z_2 \{ Q \}.
        \label{ind4}
\end{equation}
On the other hand
\begin{equation}
  \frac{\delta Z_1 \{Q\}}{\delta Q (\vr)} = \frac{Z_1 \{Q\}}{4 \tau}
  G(\vr,\vr) = \frac{\pi \nu }{4 \tau} \langle g_\vn ^{(0)}
  \rangle \cdot Z_1 \{Q\}.
        \label{ind5}
\end{equation}
Thus, the functionals $Z_{1,2} \{Q\}$ obey indentical equations. Since
$Z_1 \{\Lambda\} =Z_2 \{\Lambda\} = 1 $ the indentity (\ref {ind3}) is proven.

Substituting Eq.~(\ref{ind3}) into Eq (\ref {log1}) and taking the
Gaussian integral over $Q$, we arrive at the final
expression (\ref{BigVic}).

%%%%%%%%%%%%%%%%%%%%%%%%%%%%%%%%%%%%%%%%%%%%%%%%%%%%%%%%%%%%%%
$\bf 5$. For small gradients, the free energy (\ref {BigVic}) reduces
to the standard $\sigma$-model  (\ref {Ef}). To show this we expand
the matrix
$g_\vn$ into the sum over sperical functions
$Y_{L,M} (\vn)$
$$g_\vn (\vr) = \sum_{L=0}^{\infty}
\sum_{M=-L}^L g_{L,M} (\vr) \cdot
Y_{L,M} (\vn)$$
and note that only zero and first harmonics contribute
to the functional integral (\ref{BigVic}):
\begin{equation}
g_\vn = Q(\vr) + \vJ (\vr) \cdot \vn - \frac{Q \vJ^2}{6}.
        \label{subst}
\end{equation}
The constraint $g^2 = 1$ now reads
\begin{equation}
        Q^2 =1, \qquad Q \vJ + \vJ Q = 0.
 \label{constr7}
\end{equation}
Substituting the Eq.~(\ref {subst}) into Eqs.~(\ref {BigVic}) and using
conditions (\ref{constr7}) we obtain the partition function in the form
\begin{equation}
  Z= \int {\cal D} Q \int {\cal D} \vJ e^{-F(Q, \vJ)}, \quad F(Q, \vJ) =
  \frac{\pi \nu}{4} \int d \vr \str \{ i\omega \Lambda Q +
  \frac{\vJ^2}{6\tau } - \frac{v_F}{3} (\vnabla Q) Q \vJ \}
        \label{poldorogi}
\end{equation}
After the Gaussian integration over $\vJ$ in Eq (\ref {poldorogi}) we arrive,
finally, at Eq (\ref {Ef}).

%%%%%%%%%%%%%%%%%%%%%%%%%%%%%%%%%%%
$\bf 6$. Equations (\ref {BigVic}) can be generalised in order to describe
the ballistic motion in the presence of external fields. In a general case the
electron is described by the classical hamiltonian $H(p_i , x_i )$ and
the kinetic equation (\ref{Eilen2}) has the form (see \cite {Schmid}):
\begin{equation}
        \{H(x,p),g(x,p)\}=\left[\left(\frac{i\omega
\Lambda}{2} - \frac{Q}{2\tau} \right),g(p,x)\right]
        \label{Eilen3}
\end{equation}
where $\{H,g\}$ denotes the Poisson brackets
$$ \{H(x,p),g(x,p)\}=\frac{\partial H}{\partial p_i}\frac{\partial
  g}{\partial x_i} - \frac{\partial H}{\partial x_i}\frac{\partial
  g}{\partial p_i}
$$
Equation (\ref{Eilen3}) is still the first order
differential equation and the generalisation of
expression~(\ref{BigVic2}) for the free energy has the form
\begin{equation}
        F = \frac{\pi}{4} \int d x_i dp_i \delta (E-H(p,x))
        \str  \left\{ i\omega \Lambda g -
         \frac{g \langle g \rangle}{2 \tau}
        - \frac{1}{2} \int_0^1 du {\tilde g} (x,p, u)
        \left[ \frac{\partial {\tilde g}}{\partial u},
        \{H,{\tilde g}\} \right] \right\}
        \label{GreatVic}
\end{equation}
where
$$ \langle g(x) \rangle = \frac{1}{\nu} \int dp'_i \delta (E - H(p', x))
g(x,p').
$$

%%%%%%%%%%%%%%%%%%%%%%%%%%%%%%%%%%%%%%%%%%%%%%%%%%%%%%%%%%%%%%%%%%

$\bf 7$. As an application of Eq.~(\ref {GreatVic}), let us consider
the derivation  of the Pruisken action \cite{Pruisken}
for a two-dimensional electron gas
in a perpendicular magnetic field~$B$. To simplify the treatment, we
consider only the case of classically weak field
\begin{equation}
        \Omega_c \tau \ll 1; \qquad \Omega_c = \frac{eB}{mc},
        \label{Criter}
\end{equation}
when there is no Landau quantisation and the density of states
$\nu$ is a constant. Nevertheless, we take into account that in
the presence of magnetic field the symmetry of $g$-matrix is reduced,
and $g$ belongs to the unitary ensemble. The Poisson brackets in magnetic
field are
\begin{equation}
 \{H,g\} = v_F \vn \frac{\partial g_\vn}{\partial \vr}
 + \Omega_c \left[ \vn \times \frac{\partial g_\vn} {\partial \vn} \right]
        \label{Eilen4}
\end{equation}
and the free energy (\ref{GreatVic}) has the following form
\begin{equation}
  F = \frac{\pi \nu}{4} \int d\vr \str \left\{ i\omega \Lambda \langle g
    \rangle - \frac{\langle g \rangle ^2}{2 \tau} -\frac{1}{2} \int_0^1
    du \left \langle {\tilde g} (x,p, u) \left[\frac{\partial
          {\tilde g}}{\partial u} , v_F \vn
        \frac{\partial \tilde g}{\partial \vr } + \Omega_c \left[ \vn \times
        \frac{\partial \tilde g}{\partial \vn}
        \right] \right] \right \rangle \right\}
        \label{GreatVicMag}
\end{equation}
In the diffusive limit the expansion (\ref{subst}) can be used, which leads to
the following expression for the free energy as a functional of $Q$ and
$\vJ$;
\begin{equation}
  Z= \int {\cal D} Q \int {\cal D} \vJ e^{-F(Q, \vJ)}, \quad F(Q, \vJ) =
  \frac{\pi \nu}{4} \int d \vr \str \{ i\omega \Lambda Q + \frac{
    \vJ^2}{4\tau } - \frac{v_F}{2} (\vnabla Q)Q \vJ  - \frac{\Omega_c}{2} Q
    \left[\vJ \times \vJ \right] \}
        \label{FQJ}
\end{equation}
The last term in the free energy (\ref {FQJ}) does not vanish because the
components of the matrix $\vJ$ do not commute.
Under the conditions (\ref{Criter}), the Gaussian integration over $ \vJ $ may
be performed, with the vector product in Eq.~(\ref{FQJ})
as a perturbation, to yeild, finally, the free energy in the form
\begin{equation}
  F=\frac{\pi}{8 e^2}\int d \vr \str \left( \sigma_{xx} (\vnabla Q)^2 +
    2\sigma_{xy} Q[\nabla_x Q,\nabla_y Q] \right)
        \label{Pruis}
\end{equation}
where
\begin{equation}
\sigma_{xx} = e^2 \nu D , \qquad \sigma_{xy} = \sigma_{xx} \cdot \Omega_c \tau
        \label{sigmas}
\end{equation}

%%%%%%%%%%%%%%%%%%%%%%%%%%%%%%%%%%%%%%%%%%%%%%%%%%%%%%%%%%%%%%%%%%%%

$\bf 8$. There is a topological question, related to the
${\cal W}$-term in the free energy~(\ref{BigVic}): is it always
possible to construct the functional ${\cal W}\{g\}$, whose variation
is given by Eq.~(\ref{deltaW})? The prescription (\ref{W}) gives the
${\cal W}$-term for the functions $g(\vr)$, which are close to $g_0
(\vr) \equiv \Lambda$. The question is whether such a functional
can be defined globaly.

The answer depends upon the topology of the constant energy surface
$H(\vr,\vp)=E$ in the phase space $\{x_i , p_i \}$. For the cases of
billiards and space dimension $d > 1$ the functional ${\cal W}$ does exist.

For a one-dimensional system ${\cal W}$ can only be found as a
multivalued functional, just as the action
(\ref{Klaud}). This causes no trouble, provided $\pi \hbar \nu v_F $ is
an integer. This integer exactly equal to the wave-guide channel
number in the wire.

An accurate mathematical formulation and the proof of these statesments
will be presented elsewhere.

%%%%%%%%%%%%%%%%%%%%%%%%%%%%%%%%%%%%%%%%%%%%%%%%%%%%%%%%%%%%%%

$\bf 9.$ So far, we have considered only the systems with finite
amount of disorder. One can see, however, that the expression
(\ref{GreatVic}) remains meaningful even as $\tau \to \infty$.
Therefore, we expect that the free energy $F_\infty= F(\tau \to \infty)$
describes a clean system with the Hamiltonian $H$. As a consequence, the
partition function $Z_\infty= \int {\cal D} g \exp ( - F_\infty)$ with
the proper source terms gives the level statistics.

In the low-frequency limit ($\omega \to 0$) only the zero-mode $g^0(r,p)$
such that $\{H,g^0\}=0$ contributes to $Z_\infty$.  There are two
possibilities:
\begin{enumerate}
\item The hamiltonian system under cosideration is integrable and there
  exists a set of integrals of motion $\{ I_1, \ldots I_n \}, \quad \{H,
  I_k\}=0$. Under this condition the energy levels are characterised by
  the eigenvalues of $\{ I_1, \ldots I_n \}$ and do not repel each other.
  Therefore the level statistics is Poissonian.
\item The classical dynamics is chaotic and the only integral of motion
  is energy. In this case the zero-mode is constant in the phase space
  and $Z_\infty$ is reduced to the form
\begin{equation}
  Z_\infty= \int\limits_{g^2=1} \!\! {\cal D} g \exp \left( - \frac{\pi \nu
      \omega}{4} \str (\Lambda g) \right)
\label{zero-mode}
\end{equation}
which leads to the Wigner-Dyson (WD) level statistics \cite{Efetov}.
\end{enumerate}

In the chaotic case deviations form the WD statistics occur for the
frequencies larger than the inverse time of flight through the system.
These deviations are described by the small fluctations of $g$ about
several stationary points $\Lambda_i$, similar to what has been recently
shown by Andreev and Altshuler (AA) for diffusive systems \cite{AA}. In
complete agreement with a general AA-conjecture, the deviation from the
WD statistics is described by the determinat of some operator. It follows
from our consideration that this is the Liovillean operator
$$
\hat{L} = \frac{\partial H}{\partial p} \cdot
\frac{\partial}{\partial x} - \frac{\partial H}{\partial x} \cdot
\frac{\partial}{\partial p}
$$
%%%%%%%%%%%%%%%%%%%%%%%%%%%%%%%%%%%%%%%%%%%%%%%%%%%%%%%%%%%%%%%%%%%%%%%%%%

$\bf 10$. In conclusion, we would like to emphasize that the theory presented
here contains the diffusive $\sigma$-model as a limiting case and supplies it
with the physically motivated regularisation of the infinities
at short distances.

We are greatful to B.D.Simons for his question, which inspired
 us to derive Eq (\ref{Pruis}).

%%%%%%%%%%%%%%%%%%%%%%%%%%%%%%%%%%

\end{document}